\documentstyle[aps,prl,twocolumn,floats,graphicx]{revtex}

\begin{document}
\draft \preprint{}
\title{Photoinduced Changes of Reflectivity in Single Crystals of $YBa_{2}Cu_{3}O_{6.5}$ (Ortho II)}
\wideabs{
\author{G. P. Segre, N. Gedik, J. Orenstein}
\address{Materials Sciences Division, Lawrence Berkeley National Laboratory and Physics Department, University
of California, Berkeley, California 94720}
\author{D.A. Bonn, Ruixing Liang, W.N. Hardy}
\address{University of British Columbia, Vancouver, Canada}
\date{\today}
\maketitle
\begin{abstract}
We report measurements of the photoinduced change in reflectivity
of an untwinned single crystal of $YBa_{2}Cu_{3}O_{6.5}$ in the
ortho II structure.  The decay rate of the transient change in
reflectivity is found to decrease rapidly with decreasing
temperature and, below $T_c$, with decreasing laser intensity.  We
interpret the decay as a process of thermalization of antinodal
quasiparticles, whose rate is determined by an inelastic
scattering rate of quasiparticle pairs.
\end{abstract}
}
 Much of current researchon cuprate superconductivity focuses on the low-energy electronic
structure. Perhaps the most pressing question is whether the
pseudogap is a precursor of the superconducting gap, or has an
entirely different origin.  Another key question is whether
quasiparticles live in a two-dimensional environment, or one with
pronounced one-dimensional properties on short length and time
scales.

The relaxation dynamics of non-equilibrium quasiparticles can be
extremely sensitive to such local electronic properties
\cite{kabanov,demsar}. A promising technique to study
quasiparticle relaxation is optical pump and probe spectroscopy.
In this type of experiment a pulse of light stimulates a
non-equilibrium state with excess energy and quasiparticle
density.  A weaker pulse, which is delayed relative to the pump,
probes the nonequilibrium state by detecting changes in the
absorptance or reflectance. Dynamics can be measured with
sub-picosecond time resolution while anisotropy can be explored by
varying the polarization of the pump and probe beams.

In this paper we report pump and probe measurements performed on
an untwinned single crystal of $YBa_{2}Cu_{3}O_{6.5}$ (Ortho II).
In this material the charge reservoir layer is an ordered array of
alternating filled and empty copper oxygen chains.  Together with
$YBa_{2}Cu_{4}O_{8}$ it is one of two underdoped materials in
which doping does not introduce disorder.  We find, in contrast
with an earlier report \cite{mihailovic} on underdoped films, that
the decay rate of the photoinduced change in reflectivity depends
strongly on the excitation density and temperature. Because it
vanishes in the limit that both tend to zero, we interpret the
decay rate as an inelastic scattering rate of quasiparticle pairs.
We present evidence that the decay corresponds to a process of
thermalization, rather than recombination, of the nonequilibrium
quasiparticle population.  The dynamics of thermalization change
suddenly upon crossing from the pseudogap regime to the
superconducting state.

We measured the decay of the  change in reflectivity at a photon
energy of 1.5 eV due to excitation with photons of the same
energy.  The optical pulses, produced by a mode-locked Ti:Sapphire
laser, have duration of 100 fs, repetition rate 12 ns, and center
wavelength 800 nm.  Both the pump and probe beams were focused
onto the sample with a 20 cm focal length lens, yielding a spot
size of 75 $\mu$ diameter.  A photoelastic modulator varied the
pump intensity at 100 kHz and a vibrating mirror oscillating at 40
Hz varied the time delay between pump and probe.  This double
modulation provides sensitivity to the fractional change in
reflectivity $\Delta R/R$ of $10^{-7}$.

\begin{figure}[h]
     \includegraphics[width=3.25in]{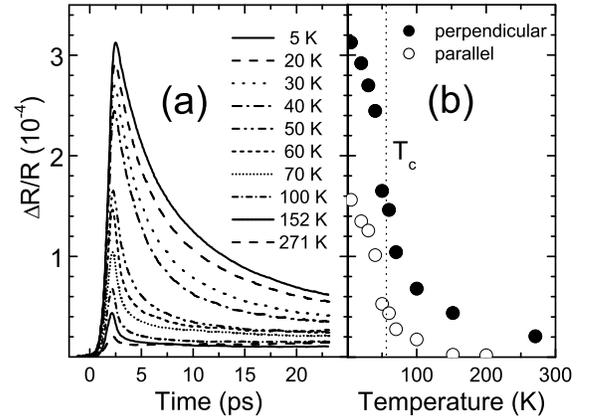}
\caption{(a) $\Delta R/R$ $vs.$ $t$ at constant pump intensity for
different temperatures (polarization perpendicular to chains). (b)
Maximum amplitude of $\Delta R/R$ $vs.$ $T$ for both polarization
parallel and perpendicular to the chains.} \label{fig:First}
\end{figure}

Fig. 1a shows $\Delta R/R$ due to absorption of a pump pulse of
energy 0.07 nJ vs. time ($t$), at several temperatures. These
reflectivity transients were measured with pump and probe
polarization each aligned perpendicular to the chain axis but the
decay rate is the same when measured with polarization along the
chains. The amplitude of $\Delta R/R$ increases and the decay rate
decreases as the temperature is lowered. Fig. 1b shows the
temperature dependence of the maximum amplitude $(\Delta
R/R)_{max}$.  The reflectivity change is smaller when measured
with pump and probe parallel to the chains, particularly above
$T_c$.

The $T$ dependence shown in Fig. 1 is consistent with earlier
reports of $\Delta R/R$ in underdoped (twinned) thin films excited
at 1.5 eV and probed at either near-infrared \cite{kabanov} or
far-infrared \cite{kaindl} frequencies.  These data have revealed
a strong correlation of $\Delta R/R$ with the strength of spin and
current fluctuations at the characteristic energy scale for
superconductivity in the cuprates. As emphasized in
Ref.\cite{kaindl}, the increase of $\Delta R/R$ with decreasing
$T$ tracks the growth of a spin resonance peak seen by neutron
scattering \cite{rossat,mook} and a gap-like feature seen in
optical conductivity \cite{orenstein,rotter}. Furthermore, $\Delta
R/R$ shows the same behavior with varying carrier concentration as
do these spectroscopic features, namely abrupt onset at $T_c$ for
near optimal doping \cite{chwalek} and a slow onset above $T_c$
for underdoped materials \cite{kabanov,kaindl}. Such
considerations suggest that 1.5 eV photons ultimately generate
low-energy excitations, despite the fact that they initially
create an electron-hole pair whose energy is much larger than the
maximum gap energy $\Delta _{0}$.

It is generally agreed that the generation of low energy
quasiparticles occurs via a rapid cascade process, in which the
parent electron-hole pair creates a large number of excitations at
much lower energy.  What has remained less clear is the mechanism
by which this nonequilibrium population of quasiparticles causes a
change in reflectivity in the visible region of the spectrum.
However, recent measurements \cite{kaindl} of the photoinduced
change in far-IR reflectivity point to a mechanism for changes in
reflectivity at 1.5 eV. Because it is crucial to understanding the
decay of $\Delta R/R$ with time, we describe this mechanism below.

The far-IR measurements reveal that photoexcitation shifts
conductivity spectral weight from low energies (most likely the
condensate contribution at zero energy) to $\sim$100 meV. This
change corresponds to filling in the gap feature in the optical
conductivity.  The Kramers-Kronig relations dictate that this
redistribution of spectral weight changes the real part of the
dielectric function at much higher frequencies $( \omega )$
according to:

\begin{equation}
\Delta \epsilon _{1} (\omega)= - \frac{8}{\omega ^{4}} \int d
\omega^{\prime}\Delta \sigma_1 (\omega^{\prime}) \omega^{\prime2}
\label{eq:first}
\end{equation}
where $\Delta \sigma _{1}$ is the change in the real part of the
optical conductivity.  Eq. 1 shows that $\Delta \epsilon _{1}$ is
proportional to the second moment of the change in the
conductivity, or $\Sigma ^{(2)}$.  Because the spectral weight
shifts from zero to $\sim$100 meV, we can relate the second moment
to the spectral weight shift, $i.e.$, $\Sigma ^{(2)} \approx (100
meV)^2\int_{0^+} d \omega \Delta \sigma _{1} (\omega)$. Finally,
using literature \cite{cooper} values $\epsilon _{1}=1.1$ and
$\epsilon _{2}=2.7$ at 1.5 eV, we can establish an approximate
calibration between $\Delta R/R$ and $\int_{0^+} d \omega \Delta
\sigma _{1} (\omega)$.

The data shown in Fig. 2 provide a test of this mechanism for
reflectivity change at 1.5 eV.  We plot $(\Delta R/R)_{max}$ at 15
K vs. excitation density expressed in units of photons per planar
Cu. The right-hand scale shows the spectral weight shift that,
according to the preceding analysis, generates the change in
reflectivity shown on the left-hand scale. The spectral weight is
normalized to the condensate spectral weight assuming a
penetration depth of 1600 \AA \cite{basov}. Notice that $\Delta
R/R$ departs from linearity at $\sim1.5\times 10^{-4}$, which
corresponds to a shift of 30$\%$ of the condensate spectral
weight. This reasonable value for the onset of nonlinearity lends
credibility to the calibration. It is important to note that a
measurable reflectivity change occurs only if spectral weight
shifts to frequencies of the order of $\Delta_0$. If instead the
spectral weight were shifted to the quasiparticle Drude peak,
$\Delta R/R$ would be smaller by the factor $(1/\tau \Delta
_{0})^2$, which is $\sim 10^6$ for these exceptionally clean ortho
II crystals \cite{microwaveonortho}.

\begin{figure}[h]
     \includegraphics[width=3.25in]{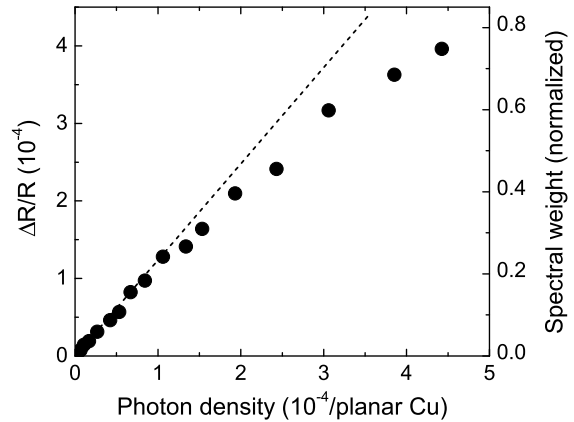}
\caption{$\Delta R/R$(black circles) $vs.$ excitation density.
Right-hand scale shows the spectral weight shift necessary to
produce the reflectivity change  on the left.} \label{fig:Second}
\end{figure}

Given this interpretation of the change in reflectivity, we now
turn to the central result of this paper, namely the rate of decay
of $\Delta R/R$ measured as a function of $T$ and pump intensity
$I$.  We have seen in Fig. 1 that the decay of $\Delta R/R$
becomes slower as $T$ decreases. Fig. 3a, a plot of the
reflectivity change at 15 K created by pulses of varying
intensity, shows that the decay rate also slows as the excitation
density decreases.

\begin{figure}[h]
     \includegraphics[width=3.0in]{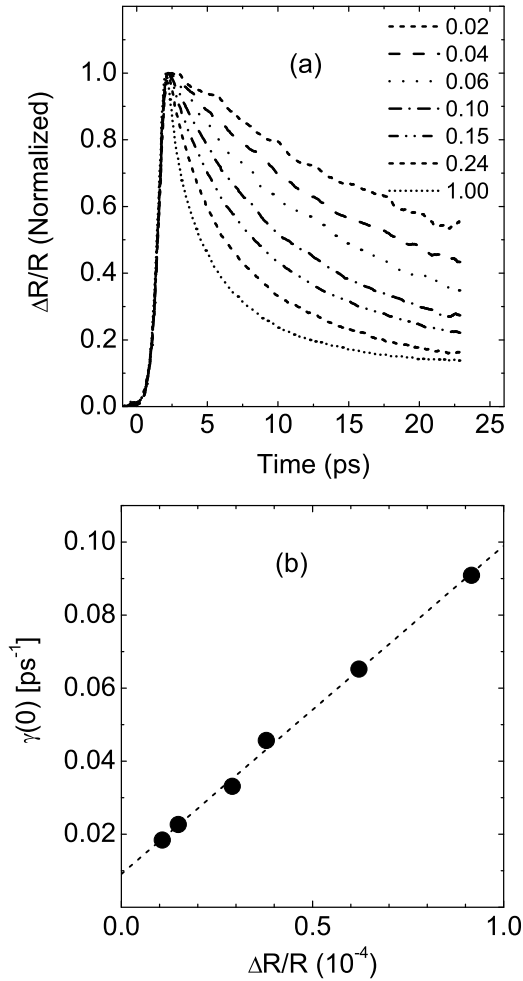}
\caption{(a) Decay of $\Delta R/R$ for several pump intensities at
15K (b)Initial decay rate $\gamma (0)$ $vs.$ $(\Delta R/R)
_{max}$.} \label{fig:Third}
\end{figure}

The decays shown in Fig. 3a are non-exponential in that the
instantaneous rate of decay, $\gamma (t)\equiv -d ln(\Delta
R)/dt$, decreases as $t$ increases.  To characterize the intensity
dependence of the decay rate in a relatively simple way, we focus
on the initial decay rate, or $\gamma (0)$.  We estimate $\gamma
(0)$ from the slope of a linear fit to $\Delta R/R$ between
100$\%$ and 75$\%$ of $(\Delta R/R) _{max}$, normalized by the
average value of $\Delta R(t)/R$ in that interval.  Given our time
resolution, this procedure provides a good estimate of $\gamma
(0)$ for rates slower than $\sim 1 ps^{-1}$. Fig. 3b shows $\gamma
(0)$ as a function of $(\Delta R/R) _{max}$, obtained by analysis
of the curves similar to the ones in Fig. 3a. $\gamma (0)$
decreases linearly with $(\Delta R/R) _{max}$ and extrapolates to
a nonzero intercept as the excitation density approaches zero.

The dependence of $\gamma (0)$ on excitation density is consistent
with two-particle kinetics, for example a process in which a pair
of quasiparticles scatter off of each other.  For high excitation
densities the rate of such scattering is proportional to the
nonequilibrium density.  As the excitation density decreases,
becoming smaller than the thermal population, $\gamma (0)$
approaches the equilibrium scattering rate at that temperature
\cite{gray}.

Fig. 4 presents the intensity dependence of the decay rate for a
range of temperatures, in the form of a logarithmic plot of
$\gamma (0)$ vs. $T$.  Points with the same (open) symbol show
$\gamma (0)$ measured at the same excitation density, but at
different temperatures.  The data sets coincide at high
temperatures but diverge suddenly at $T_c$, showing that the
dependence of $\gamma (0)$ on $I$ starts abruptly upon crossing
from the pseudogap to the superconducting state. The solid symbols
show the decay rate in the limit of zero excitation density,
$\gamma_{th}$, as determined from the intercept of plots such as
in Fig. 3b. The same quantity is plotted on linear axes in the
inset. With decreasing temperature $\gamma _{th}$ begins to
decrease in the normal state and is proportional to $T^3$ for
temperatures less than $\sim$40 K.

\begin{figure}[h]
     \includegraphics[width=3.25in]{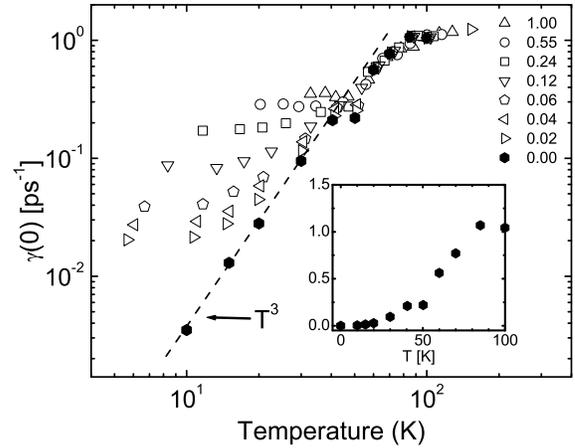}
\caption{Initial decay rate $\gamma(0)$ $vs.$ $T$ at fixed $I$
(open symbols). Decay rate in low $I$ limit, $\gamma_{th}$, $vs.$
$T$ (solid symbols in main panel and inset).} \label{fig:Fourth}
\end{figure}

Several aspects of the decay rate of $\Delta R/R$ suggest a strong
connection to quasiparticle dynamics, as well as structure in the
low-energy density of states.  For example, the decay rate appears
sensitive to the opening of the pseudogap, decreasing near $T^*$
in a manner which is remarkably similar to the nuclear spin
relaxation rate in underdoped YBCO \cite{warren}.  The onset of
superconductivity also affects the decay dynamics --- the decay
rate depends on excitation density only below $T_c$.  Finally the
$T^3$ dependence of $\gamma_{th}$ is consistent with the
prediction of a theoretical analysis \cite{quinlan} of
quasiparticle scattering and recombination in the presence of
antiferromagnetic correlations. Although the rates were calculated
within the random phase approximation, the asymptotic $T^3$
behavior is a universal feature of quasiparticle scattering in a
d-wave superconductor \cite{quinlan}.

The simplest interpretation of the connection between $\gamma(0)$
and quasiparticle dynamics is that $\Delta R(t)/R$ reflects the
decay of the nonequilibrium quasiparticle density. According to
this view, $\gamma(0)$ is the rate at which quasiparticle pairs
scatter into the condensate and $\gamma_{th}$ is its value in
thermal equilibrium. Although such an interpretation appears to be
consistent with the prediction of Ref. 16, this description of the
decay of $\Delta R/R$ is not internally consistent.  The
inconsistency arises because quasiparticle interactions alone
cannot reduce either the excess particle or energy density. For
quasiparticles to scatter into the condensate they must shed their
excess energy, which for antiferromagnetic interactions appears as
a spin fluctuation. However, in the low-energy regime
($\omega<<J$) the spin fluctuation is essentially the same entity
as a quasiparticle-hole pair. Therefore such a 'recombination'
process actually scatters one pair into another, with no net
change in the number or energy density. This is analogous to a
problem in the interpretation of transport experiments, where
quasiparticle scattering alone cannot reduce the total momentum
\cite{walker}.

The 'true' recombination rate, in the sense of recovery of the of
equilibrium  quasiparticle density, would likely differ markedly
from the reflectivity decay rates shown in Fig. 4.  Relaxation of
the quasiparticle density requires that excess energy flow
irreversibly from the electrons to the lattice \cite{feenstra}. We
expect that process to be extremely slow on the time scale of
quasiparticle scattering and to exhibit single rather than
two-particle kinetics. This expectation is based on the physics of
the 'phonon bottleneck' in s-wave superconductors\cite{gray}, an
effect which may be even more restrictive in d-wave
superconductors.

The bottleneck arises because the process in which a quasiparticle
pair emits a phonon and scatters into the condensate is reversible
--- the emitted phonon can immediately regenerate the
quasiparticle pair. The quasiparticle density relaxes at the rate
at which the typical phonon energy decreases below the threshold
for pair creation. Not only is this rate slower than the
quasiparticle scattering rate, it is also approximately
independent of the excess quasiparticle density.

The relaxation of the quasiparticle density may be even slower in
the cuprates than in conventional superconductors.  In s-wave
superconductors the phonons decouple from the electrons when their
energy becomes less than $2\Delta$. In the d-wave superconductors,
where there is no threshold for pair creation, the typical phonon
energy must degrade to $\sim k_{B}T$ in order for the electronic
system to reach equilibrium.  The extremely weak coupling between
nodal quasiparticles and phonons compounds this effect.  The decay
of a quasiparticle pair into an acoustic phonon is kinematically
forbidden because the quasiparticle velocity is larger than the
sound velocity \cite{feenstra}.

If recombination is extremely slow, why does $\Delta R/R$ decay on
a picosecond time scale? The remaining possibility is that the
decay corresponds to thermalization, rather than recombination. If
$\Delta R/R$ is not a direct measure of the quasiparticle density,
but proportional instead to the $\Sigma^{(2)}$ as we have argued
previously, it can decrease during thermalization, despite a
constant, or even increasing, density of quasiparticles. Microwave
measurements show that $\Sigma^{(2)}$ for nodal particles is
small, because their spectral weight is confined to a very narrow
($\sim$10 $\mu$eV) Drude peak \cite{hosseini}. The observed
$\sim$100 meV spectral weight shifts \cite{kaindl}, and
consequently large values of $\Sigma^{(2)}$, are likely due to the
antinodal quasiparticles. These considerations suggest that the
decay of $\Delta R/R$ measured at 1.5 eV tracks the conversion of
quasiparticles from a nonthermal distribution, which has
substantial $\Sigma^{(2)}$, to a quasithermal distribution in
which $\Sigma^{(2)}$ is a factor $10^6$ smaller. Measurements in
which the nonequilibrium state is probed in the frequency range of
the Drude conductivity \cite{feenstra} show much longer decay
times (typically milliseconds), consistent with an experiment
which selectively probes the nodal quasiparticle density.

In summary, we present measurements of the photoinduced change in
reflectivity in an underdoped crystal of YBCO, and discuss a new
perspective on this effect. We interpret the decay rate of $\Delta
R/R$ as the inelastic scattering rate of quasiparticle pairs,
which sets the time scale for thermalization of the nonequilibrium
population. The thermalization rate decreases with the development
of the pseudogap, becomes excitation density dependent in the
superconducting state, and varies as $T^3$ in the low excitation
limit. Coordinated measurements of photoinduced changes in optical
response over a broad spectral range, from terahertz to visible
frequencies, are needed to test this perspective.

We acknowledge valuable discussions with D.J. Scalapino.  This
work was supported under NSF-9870258, DOE-DE-AC03-76SF00098,
Canadian Institute for Advanced Research and NSERC.

\end{document}